\documentclass[mathleft,fleqn,%
]{an}
\usepackage{graphicx}
\usepackage[varg]{txfonts}
\begin{document}

\Pagespan{1}{}
\Yearpublication{2014}%
\Yearsubmission{2014}%
\Month{0}%
\Volume{999}%
\Issue{0}%
\DOI{asna.201400000}%

\title{Milky Way chemo-dynamics in the era of Gaia}


\author{I.\ Minchev\inst{1}\fnmsep\thanks{Corresponding author:
        {iminchev@aip.de}}
\and C.\ Chiappini\inst{1}
\and M.\ Martig\inst{2}
}
\titlerunning{Milky Way chemo-dynamics}
\authorrunning{Minchev et al.}
\institute{
Leibniz-Institut f\"{ur} Astrophysik Potsdam (AIP), An der Sternwarte 16, D-14482, Potsdam, Germany
\and 
Max-Planck-Institut f\"{ur} Astronomie, K\"{o}nigstuhl 17, D-69117 Heidelberg, Germany
}
\received{XXXX}
\accepted{XXXX}
\publonline{XXXX}


\abstract{%
The main goal of Galactic Archaeology is understanding the formation and evolution of the basic Galactic components. This requires sophisticated chemo-dynamical modeling, where disc asymmetries (e.g., perturbations from the central bar, spirals arms, and infalling satellites) and non-equilibrium processes are taken into account self-consistently. Here we discuss the current status of Galactic chemo-dynamical modeling and focus on a recent hybrid technique, which helps circumvent traditional problems with chemical enrichment and star formation encountered in fully self-consistent cosmological simulations. We show that this model can account for a number of chemo-kinematic relations in the Milky Way. In addition, we demonstrate that (1) our model matches well the observed age-[$\alpha$/Fe] and age-[Fe/H] relations and (2) that the scatter in the age-[Fe/H] relation cannot be simply explained by blurring (stars on apo- and pericenters visiting the solar vicinity) but significant radial migration (stars born elsewhere but ending up at the solar vicinity today because of a change in guiding radius) is needed. 
We emphasize the importance of accurate stellar ages, such as those obtained through asteroseismology by the CoRoT and Kepler missions, for breaking the degeneracy among different Galactic evolution scenarios. 
}
\maketitle

\section{Introduction}
Important information concerning the dominant mechanisms responsible for the formation of the Milky Way (MW) is encoded in the chemistry and kinematics of its stars. A number of Galactic surveys are currently being conducted with the aim of obtaining chemical and kinematical information for a vast number of stars, e.g., RAVE (Steinmetz et a. 2006), SEGUE (Yanny et a. 2009), APOGEE (Majewski et al. 2010), HERMES (Freeman et al. 2010), Gaia-ESO (Gilmore et al. 2012), Gaia (Perryman et al. 2001), LAMOST (Zhao et al. 2006) and 4MOST (de Jong et al. 2012). 

To be able to interpret the large amounts of forthcoming data we need galaxy formation models tailored to the MW. Producing disc-dominated galaxies has traditionally been a challenge for cosmological simulations (e.g., Navarro et al. 1991; Abadi et al. 2003). Although an increase in resolution and improved modeling of star formation and feedback have resulted in MW-mass galaxies with reduced bulge fractions (Guedes et al. 2011; Martig et al. 2012), typically these simulations do not include chemical evolution. Galaxy formation simulations including some treatment of chemical evolution have been performed by several groups (e.g., Scannapieco et al. 2005; Few et al. 2012). 

However, while the results are encouraging and global observed relations are reproduced, such as the metallicity trends between the different galactic components (e.g.,Tissera et al. 2012), it is still a challenge for cosmological simulations to match the properties of the MW (e.g., the typical metallicities of the different components -- Tissera et al. 2012). Additionally, the fraction of low-metallicity stars are often overestimated (Calura et al. 2012) and reproducing the position of thin- and thick-disc stars in the [$\alpha$/Fe]-[Fe/H] plane has proved challenging (Brook et al. 2012; Gibson et al. 2013). While such issues could be due to unresolved metal mixing (Wiersma et al. 2009), none of the above-mentioned simulations reproduces simultaneously the mass, the morphology, and the star formation history (SFH) of the MW.

\subsection{Summary of MW chemo-dynamical evolution modeling techniques}

A major consideration in a disc chemo-dynamical model is taking into account the effect of radial migration, i.e., the fact that stars end up away from their birth places (see, e.g., Sellwood and Binney 2002; Roskar et al. 2008, Minchev and Famaey 2010). Below we quickly summarize models which include radial migration.

1. Semi-analytical models tuned to fit the local metallicity distribution, velocity dispersion, and chemical gradients, etc., today (e.g., Schoenrich and Binney 2009; Kubryk, Prantzos, and Athanssoula 2014) or Extended distribution functions (Sanders and Binney 2015): 
\newline$-$ Easy to vary parameters
\newline$-$ Provide good description of the disc chemo-kinematic state today
\newline$-$ Typically not concerned with the Milky Way past history
\newline$-$ Time and spatial variations of migration efficiency due to dynamics resulting from non-axisymmetric disc structure is hard to take into account.

2. Fully self-consistent cosmological simulations (e.g., Kawata and Gibson 2003; Scannapieco et al. 2005; Kobayashi and Nakasato 2011; Brook et al. 2012):
\newline$-$ Dynamics self-consistent in a cosmological context
\newline$-$ Can learn about disc formation and evolution
\newline$-$ Not much control over final chemo-kinematic state
\newline$-$ Problems with SFH and chemical enrichment due to unknown subgrid physics
\newline$-$ Much larger computational times needed if chemical enrichment included.

3. Hybrid technique using simulation in a cosmological context + a classical (semi-analytical) chemical evolution model (Minchev, Chiappini, and Martig 2013, 2014):
\newline$-$ Avoids problems with SFH and chemical enrichment in fully self-consistent models
\newline$-$ Can learn about disc formation and evolution
\newline$-$ Not easy to get Milky Way-like final states.

For the rest of the paper we focus on the results of the latter model.

\section{Our chemo-dynamical model}

To properly model the MW it is crucial to be consistent with some observational constraints at redshift $z=0$, for example, a flat rotation curve, a small bulge, a central bar of an intermediate size, gas to total disc mass ratio of $\sim0.14$ at the solar vicinity, and local disc velocity dispersions close to the observed ones.

It is clear that cosmological simulations would be the natural framework for a state-of-the-art chemodynamical study of the MW. Unfortunately, as described above, a number of star formation and chemical enrichment problems still exist in fully self-consistent simulations. We have, therefore, resorted to the next best thing -- a high-resolution simulation in the cosmological context coupled with a pure chemical evolution model (see detailed description in Minchev, Chiappini, and Martig 2013, hereafter MCM13).

The simulation we used is part of a suite of numerical experiments first presented by Martig et al. (2012), where the authors studied the evolution of 33 simulated galaxies from $z=5$ to $z=0$ using the zoom-in technique described by Martig et al. (2009). This technique consists of extracting merger and accretion histories (and geometry) for a given halo in a $\Lambda$-CDM cosmological simulation, and then re-simulating these histories at much higher resolution (150~pc spatial, and 10$^{4-5}$~M$_{\odot}$ mass resolution). The interested reader is referred to Martig et al. (2012) for more information on the simulation method. 

Originally, our galaxy has a rotational velocity at the solar radius of 210 km/s and a scale-length of $\sim5$~kpc. To match the MW in terms of dynamics, at the end of the simulation we downscale the disc radius by a factor of 1.67 and adjust the rotational velocity at the solar radius to be 220 km/s, which affects the mass of each particle according to the relation $GM\sim v^2r$, where $G$ is the gravitational constant. This places the bar's corotation resonance (CR) and 2:1 outer Lindblad resonance (OLR) at $\sim4.7$ and $\sim7.5$~kpc, respectively, consistent with a number of studies (e.g., Dehnen 2000, Minchev, Nordhaus, and Quillen 2007, Minchev et al. 2010). At the same time the disc scale-length, measured from particles of all ages in the range $3<r<15$~kpc, becomes $\sim3$~kpc, in close agreement with expectations in the MW. 

The chemical evolution model we use here is for the \emph{thin disc only}. The idea behind this is to test, once radial mixing and merger perturbations are taken into account, if we can explain the observations of both thin and thick discs without the need of invoking a discrete thick-disc component. A detailed description of the chemical model is given in MCM13.

\begin{figure*}
\includegraphics[width=17. cm]{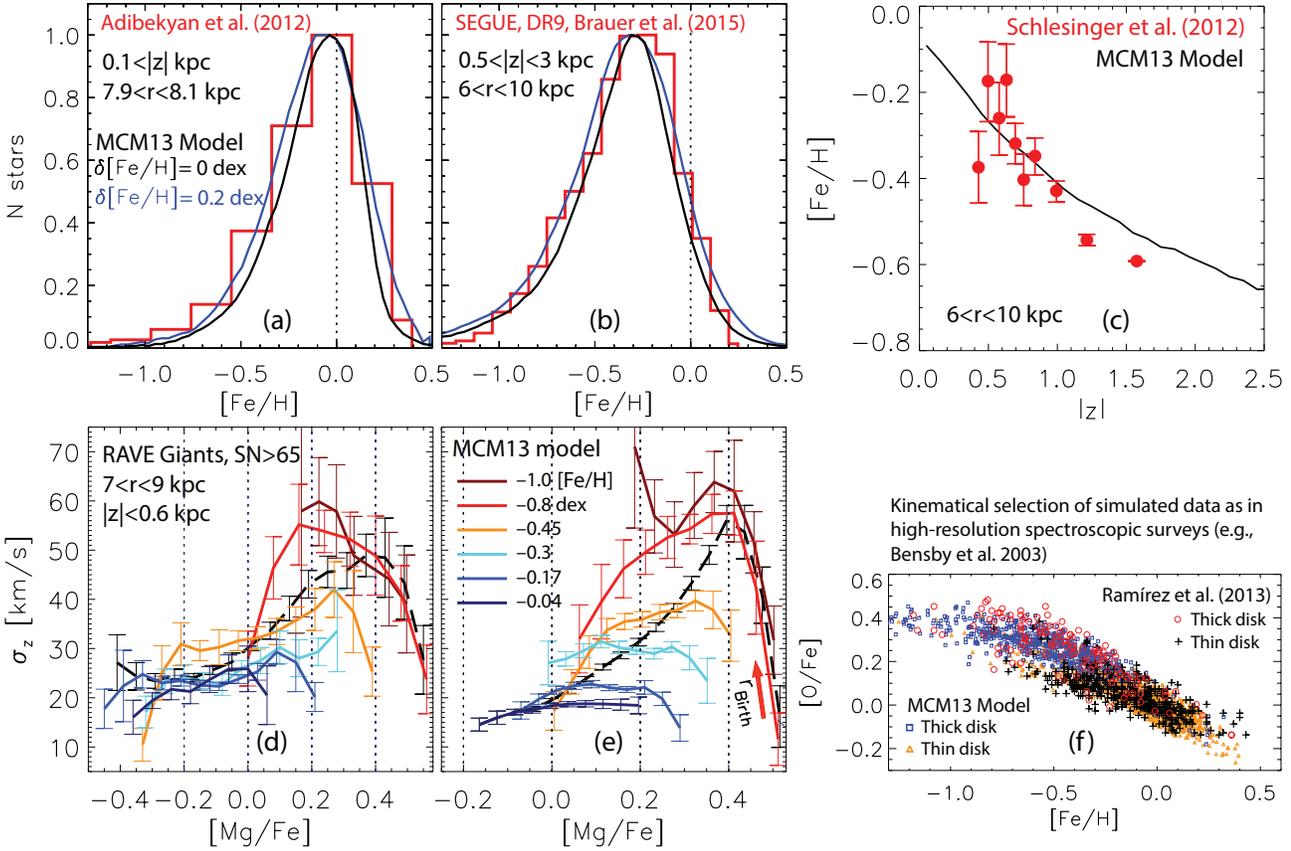}
\caption{
Comparison between the prediction of the MCM13 model and observations. 
{\bf Panels (a) and (b):} The red histograms show data from Adibekyan et al. (2012) and Bauer et al. (in preparation); the black and blue curves show the model with and without convolved error, respectively. The metallicity peak shifts to lower [Fe/H] for both the data and model, when distance from the disc plane increases. 
{\bf Panel (c):} Metallicity variation with distance from the disc plane for SEGUE G-dwarf data (red). 
{\bf Panels (d) and (e):} Variations of vertical velocity dispersion with [Mg/Fe] for RAVE giants (d) and for the model (e). 
{\bf Panel (f):} Comparison to the high-resolution data by Ramirez et al. (2013). A shift in the model [O/Fe] of 0.05 dex (within the uncertainty) has been applied. 
Panels (a), (b), (c), (f) are from MCM13 and panels (d) and (e) are from Minchev et al. (2014a).}
\label{fig:chem}      
\end{figure*}

\begin{figure*}
\centering
\includegraphics[width=17cm]{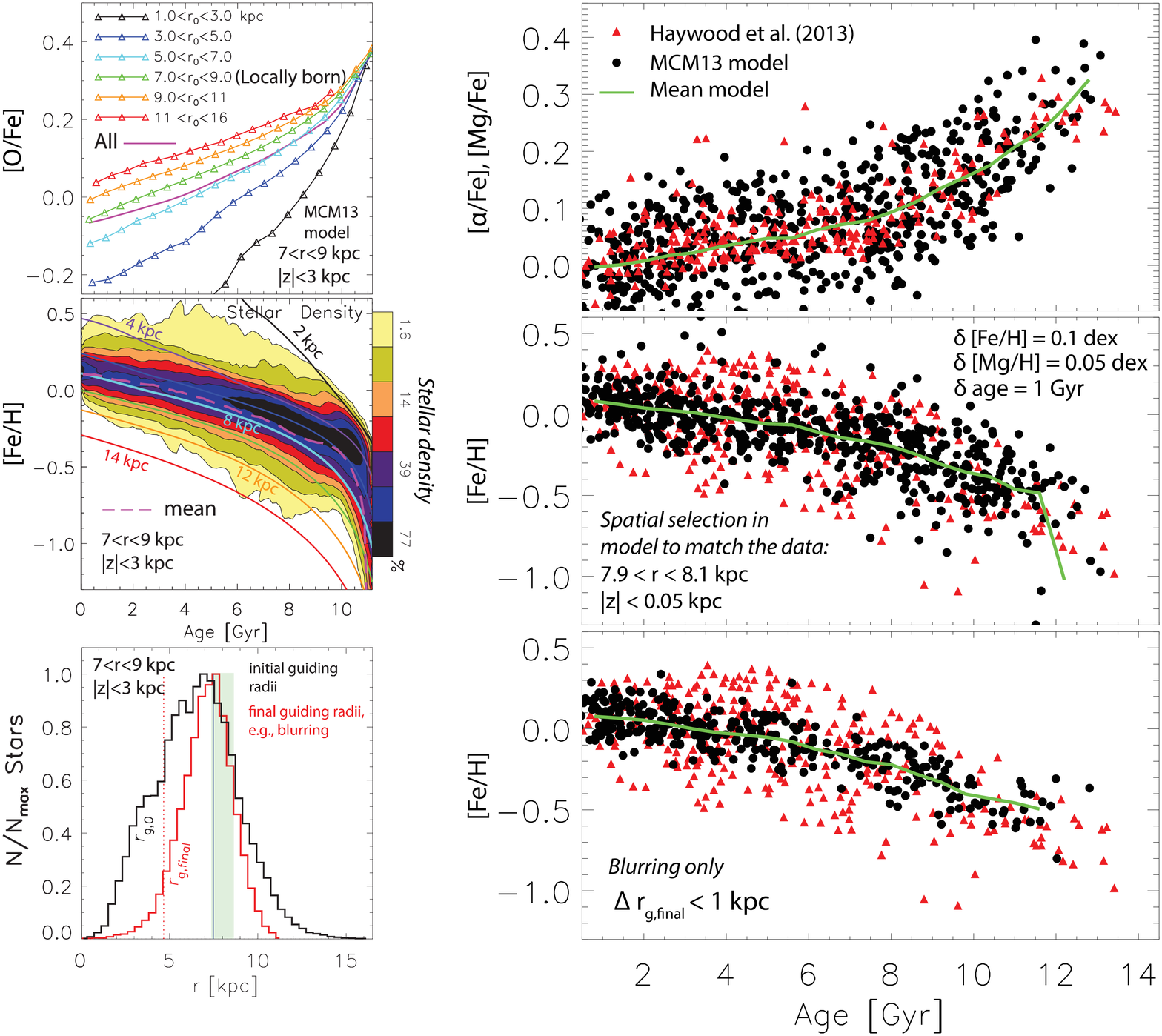}
\caption{ 
Comparison between the Haywood et al. (2013) data and MCM13 model. 
{\bf Left column:} The model age-[O/Fe] relation for stars arriving to the solar vicinity from different radii (top), the age-[Fe/H] relation (middle), and distributions of birth and final guiding radii (bottom) of stars ending up in the solar vicinity, indicated by the green strip. The dotted red and solid blue vertical lines in the bottom left panel indicate the bar's corotation and inner Lindblad resonance.
{\bf Right column:} The age-[$\alpha$/Fe] relation (top), where [Mg/Fe] from the model is used; the age-[Fe/H] relation (AMR) shows a smaller scatter in the model than in the data (middle); and the AMR but with migrators removed from the model (bottom). This figure shows that the MCM13 model is consistent with the observed relations and suggests that blurring only (i.e., lack of migration) is insufficient to explain to observed AMR. 
\label{fig:hay}
}
\end{figure*}

\subsection{Comparison to recent observational results}

The MCM13 model has been shown in previous works to match simultaneously the following observational constraints, some of which are reproduced in Fig.~\ref{fig:chem}:
\newline
\newline$\bullet$ The disc morphology, scale-length and rotation curve today (see MCM13);
\newline$\bullet$ The local age-velocity dispersion relation (Sharma et al. 2014);
\newline$\bullet$ The more centrally concentrated [$\alpha$/Fe]-enhanced (old) disc (Bensby et al. 2011; Bovy et al. 2012; see Minchev, Chiappini, and Martig 2014b, Fig.11);
\newline$\bullet$ The distribution of scale-heights for mono-abundance subpopulations found in SEGUE G-dwarfs (MCM13, Fig.13);
\newline$\bullet$ The reversal of the radial [$\alpha$/Fe] and metallicity gradients (e.g., APOGEE - Anders et al. 2014), when sample distance from the disc mid-plane is increased (Minchev et al. 2014b).
\newline$\bullet$ The MDF for stellar samples at different distance from the disc midplane (Fig.~\ref{fig:chem}, a, b). 
\newline$\bullet$ The metallicity variation with vertical distance from the plane (Fig.~\ref{fig:chem}, c);
\newline$\bullet$ The inversion in velocity dispersion relation in RAVE and SEGUE (Fig.~\ref{fig:chem}, d, e; Minchev et al. 2014; Guiglion et al. 2015);
\newline$\bullet$ The [$\alpha$/Fe]-[Fe/H] plane (current Fig.~\ref{fig:chem}, f; Ramirez et al. 2013);
\newline$\bullet$ The flaring of mono-abundance populations (assuming similarity to mono-age populations) found by Bovy et al. (2015) and predicted earlier by Minchev et al. (2015), where Model 1 in the latter paper presents the same galaxy as the one used for the MCM13 chemodynamical model.

\subsection{The age-[$\alpha$/Fe] relation}

We now compare the age-[$\alpha$/Fe] and AMR relations in the model to those resulting from the data by Haywood et al. (2013). These authors published ages for single stars with known [Fe/H] and [$\alpha$/Fe] in the immediate solar neighborhood. Their sample is based on the HARPS GTO observations of 1111 stars, described by Adibekyan et al. (2012). The original sample had to be severely reduced to 363 stars. This down-selection was based on an absolute magnitude cut at M$_{V} < 4.75$ and on a somewhat less reproducible selection of stars with "a well defined probability function". Haywood et al. noted that their absolute age scale could be off by 1 to 1.5 Gyr, while relative ages would have uncertainties of 1 Gyr. Their definition of [$\alpha$/Fe] includes the mean of Mg, Si, and Ti abundances. 

The top-left panel of Fig.~\ref{fig:hay} shows the model age-[O/Fe] relation for stars in the solar vicinity at the final time, but with different birth radii (from MCM13, Fig.~7). It is important to realize that, as seen in the bottom-left panel, the majority of stars in the solar vicinity at the final simulation time have been born close to the solar radius. The second row, left panel indicates that the scatter in the model AMR is related to stars born inside and outside the solar radius. Most stars are contained in the three highest levels (black, blue, and purple), i.e., scatter is not as large as it would be naively inferred from this particular plot. 

The top-right panel of Fig.~\ref{fig:hay} shows a comparison between the age-[$\alpha$/Fe] relation by Haywood et al. (2013) and the age-[Mg/Fe] relation resulting from our model. A small volume sample around the simulated solar radius was used, as indicated, similarly to the data. We expanded the model age range by 10\%, in order to match the data range. We also shifted the model by 0.03 dex to lower [$\alpha$/Fe] values. Uncertainties of $\delta$[Mg/H]$=0.05$ and $\delta$[Fe/H]$=0.1$~dex were implemented in the model. This results in an uncertainty of 0.11 dex in [Mg/Fe]. We also implemented age errors of $\sim1$~Gyr, similar to the relative age uncertainty quoted by Haywood et al. It can be seen that the model mean (green curve) provides a very good match to the data, while the scatter in the model is somewhat larger for this particular plot.

The second row, right panel of Fig.~\ref{fig:hay} shows the AMR, where it can be seen that the scatter is smaller in the model than in the data (contrary to what was stated by Haywood et al. 2013). Note that the data selection contains 625 stars, or about 2 times more than the observations. The obviously much smaller scatter in the model AMR is then clearly real and not a product of lower number statistics compared to the data.

Finally, the bottom-right panel of Fig.~\ref{fig:hay} shows again the model AMR, but with migrators removed by excluding stars which have changed guiding radii since their birth by more than 1 kpc. This assumption is very conservative in that the cut still leaves stars which have migrated by 1 kpc. Our selection preserves stars with hot kinematics in the sample, e.g., blurring (stars on apo- and pericenters visiting the solar vicinity) still contaminates the local metallicity distribution. We tested this by making sure that the age-velocity dispersion relation resulting from the original model selection (middle right panel of Fig.~\ref{fig:hay}) and the sample with the migrators removed, were very similar.

The very small scatter found in the bottom right panel of Fig.~\ref{fig:hay} suggests that blurring is insufficient to explain the observed AMR. In fact, as evident from the second-row left-panel of the same figure, even the significant migration present in the MCM13 model is insufficient to explain the scatter found in the data. Assuming we can trust the data presented here, we require either a model with more migration, or some other way of increasing the scatter in the AMR.

This above comparison shows that the MCM13 model is also consistent with the data by Haywood et al. (2013), contrary to the statement made in the latter paper. It also warns that a meaningful comparison between observational data and models must include the same way of presentation, e.g., in this case scatter plots of the model sample, spatially constrained as in the observations. 

\section{Conclusions}

Despite the recent advances in the field of galaxy formation and evolution, currently no self-consistent simulations exist that have the level of chemical implementation required to make detailed predictions for the number of ongoing and planned MW observational campaigns. Even in high-resolution simulations one particle represents $\sim10^4-10^5$ solar masses, which requires a number of approximations to compute this "sub-grid" physics. Here, instead, we have assumed that each particle is one star\footnote{Dynamically, this is a good assumption, since the stellar dynamics is collisionless.} and have implemented the exact SFH and chemical enrichment from a typical chemical evolution model into a state-of-the-art simulation of the formation of a galactic disc. Note that this is the first time that a chemo-dynamical model has the extra constraint of defining a realistic solar vicinity also in terms of dynamics.

Availability of accurate ages is very important to make progress in the field of Galactic Archaeology. This has recently become evident with the unexpected results of Chiappini et al. (2015) and Martig et al. (2015), who used CoRoT and Kepler asteroseismic ages, respectively, combined with APOGEE chemical information, to show the existence of significantly young high-[$\alpha$/Fe] stars. This is a largely unexpected result for chemical evolution modeling, where [$\alpha$/Fe] has been thought to always be a good proxy for age. Stellar ages for much larger samples and broader disc coverage are expected in the very near future from Kepler-2 and the Gaia mission. These will help break degeneracies and refine chemo-dynamical models, thus bringing us a step closer to understanding the formation of our Galaxy.

%

\end{document}